\documentclass[conference]{IEEEtran}
\ifCLASSINFOpdf
   \usepackage[pdftex]{graphicx}
  \graphicspath{{../pdf/}{../jpeg/}}
\else
\fi
\usepackage{amsmath}
\usepackage{todonotes}
\usepackage{listings}

\usepackage[T1]{fontenc}
\usepackage{multirow}
\usepackage{suffix}
\usepackage{url}
 \usepackage{acronym}
\usepackage{booktabs}
\usepackage{empheq}
\usepackage{xcolor}
\usepackage{enumitem}
\usepackage{threeparttable}
\usepackage{array}
\usepackage{comment}
\usepackage{hyperref}
\usepackage{amssymb,amsmath} 
\usepackage{enumerate} 
\usepackage{graphicx}
\usepackage{url}
\usepackage{wasysym}
\usepackage{multirow}
\usepackage{balance}

\usepackage{cite}
\usepackage[numbers]{natbib}     

\usepackage{expl3}
\ExplSyntaxOn
\newcommand\latinabbrev[1]{
  \peek_meaning:NTF . {
    #1\@}%
  { \peek_catcode:NTF a {
      #1., \@ }%
    {#1., \@}}}
\ExplSyntaxOff

\definecolor{lightpurple}{rgb}{0.8,0.8,1}
\definecolor{codebg}{RGB}{255,255,255}
\definecolor{commentcolor}{RGB}{11,140,11}
\newsavebox{\supbox}
\newcommand{\bsup}{\begin{lrbox}{\supbox}$\tt\scriptstyle}
\newcommand{\esup}{$\end{lrbox}{}^{\usebox{\supbox}}}

\newcolumntype{L}[1]{>{\raggedright\let\newline\\\arraybackslash\hspace{0pt}}m{#1}}
\newcolumntype{C}[1]{>{\centering\let\newline\\\arraybackslash\hspace{0pt}}m{#1}}
\newcolumntype{R}[1]{>{\raggedleft\let\newline\\\arraybackslash\hspace{0pt}}m{#1}}

\begin{document}
%
\title{The Vision of Software Clone Management: \\Past, Present, and Future (Keynote Paper) \vspace{-0.2cm}}


\author{\IEEEauthorblockN{Chanchal K. Roy~~~~~~~~~~ Minhaz F. Zibran~~~~~~~~~~~~~~Rainer Koschke$\dagger$}
\IEEEauthorblockA{University of Saskatchewan, Canada~~~~~~~~~~~~~~~~~~~~~~~~~ $\dagger$University of Bremen, Germany\\
\{chanchal.roy, minhaz.zibran\}@usask.ca, ~~~~~~~~~~~~~~koschke@informatik.uni-bremen.de}
\vspace{-0.2cm}}



%


\maketitle


\begin{abstract}

Duplicated code or code clones are a kind of code smell that have both positive and negative impacts on the development and maintenance of software systems. Software clone research in the past mostly focused on the detection and analysis of code clones, while research in recent years extends to the whole spectrum of clone management.  In the last decade, three surveys appeared in the literature, which cover the detection, analysis, and evolutionary characteristics of code clones. This paper presents a comprehensive survey on the state of the art in clone management, with in-depth investigation of clone management activities (e.g., tracing, refactoring, cost-benefit analysis) beyond the detection and analysis. This is the first survey on clone management, where we point to the achievements so far, and reveal avenues for further research necessary towards an integrated clone management system. We believe that we have done a good job in surveying the area of clone management and that this work may serve as a roadmap for future research in the area


\end{abstract}


\IEEEpeerreviewmaketitle

\section{Introduction and Motivation}\label{intro}



Copying existing code and pasting it in somewhere else followed by minor or major edits is a common practice that developers adopt to increase productivity. Such a reuse mechanism typically results in duplicate or very similar code fragments residing in the code base. Those duplicate or near-duplicate code segments are commonly known as code clones. There are many reasons why developers intentionally perform such code cloning. Obvious reasons include reuse of existing implementations without ``re-inventing the wheel". More comprehensive discussions on the reasons for code cloning can be found elsewhere~\cite{Roy_2007_Survey}.
Code clones may also appear in the code base without the awareness of the developers. Such unintentional/accidental clones may be introduced, for example, due to the use of certain design patterns,  use of certain APIs to accomplish similar programming tasks, or coding conventions imposed by the organization.

The reuse mechanism by code cloning offers some benefits. For instance, cloning of existing code that is already known to be flawless, might save the developers from probable mistakes they might have made if they had to implement the same from scratch. It also saves time and effort in devising the logic and typing the corresponding textual code. Code cloning may also help in decoupling classes or components and facilitate independent evolution of similar feature implementations.  

On the other end of the spectrum, code clones may also be detrimental in many cases. Obviously, redundant code may inflate the code base and may increase resource requirements. This may be crucial for embedded systems and systems such as hand held devices, telecommunication switches, and small sensor systems. Moreover, cloning a code snippet that contains any unknown fault may result in propagation of that fault to all copies of the faulty fragment. From the maintenance perspective, a change in one code segment may necessitate consistent changes in all clones of that fragment. Any inconsistency may introduce bugs or vulnerabilities in the system. Fowler et al.~\cite{Fowler_1999_Refactor} recognize code clones as a serious kind of code smell.

However, during the software development process, duplication cannot be avoided at times. For example, duplication may be enforced by the limitation of the programming language's necessary mechanism to implement an efficient generic solution of a problem at hand. Code generators may also generate duplicated code that the developers may have to modify. 
 
Although controversial, previous research reports empirical evidences that a significant portion (generally 9\%-17\%~\cite{Zibran_2011_CloneForecast}) of a typical software system consists of cloned code, and the proportion of code clones in the code base may be as low as 5\%~\cite{Roy_2007_Survey} and as high as even 50\%~\cite{Rieger_2004_CodeDuplicate}. Indeed, due to the negative impact of code clones in the maintenance effort, one might want to remove code clones by active refactoring, wherever feasible. However, in reality, aggressive refactoring of code clones appears not to be a very good idea~\cite{Cordy_2003_Adoption}, and not all clones are really removable through refactoring. Due to the dual role of code clones in the development and maintenance of software systems, as well as the pragmatic difficulty in avoiding or removing those, researchers and practitioners have agreed that code clones should be detected and managed efficiently~\cite{JSync,Zibran_2012_CloneSearch}.


Since the emergence of software clones as a research area in early 1990s, significant contributions over years made the field grow and become quite a mature area of research. Over the entire course of software clone research there have been only two notable general surveys on clones. Koschke~\cite{Koschke_2006_Survey}, in 2007, presented a brief summary of the important findings about different aspects of software clones including cause-effect of cloning, clone avoidance, detection, and evolution along with a set of open questions. In the same year, Roy and Cordy~\cite{Roy_2007_Survey} also published another survey containing a thorough review on those same areas with specific focus on clone detection tools and techniques. A few recent surveys either focus on detection \cite{Rattan20131165, Roy_2009_ToolEvaluation} or evolution of clones~\cite{Pate_2011_EvolutionSurvey}. In this vision paper, we provide an extensive survey on code clone research with strong emphasis on clone management and point readers to future research directions.

This paper is organized as follows. In Section~\ref{sec:LiteratureReview}, we present a systematic review on a repository of 353 publications appeared over 20 years. The review draws a ``birds-eye" view on the overall contributions and growth along different dimensions of software clone research. This survey is the outcome of careful investigation of literature beyond the said repository (described in Section~\ref{sec:LiteratureReview}), and through analysis in the light of our experience. Section~\ref{sec:Management} introduces different aspects of clone management activities starting with the definition and types of clones. While in Section~\ref{sec:Detection}, we list different stand-alone clone detection techniques, we discuss the IDE-based clone detectors in Section \ref{sec:IntegratedDetect}. We then talk about clone documentation in Section~\ref{sec:Documentation} and that of tracking over evolution in Section~\ref{sec:Tracking}. We discuss clone evolution studies including visualization of clone evolution in Section \ref{sec:Evolution}. Then, in Section~\ref{sec:Annotation}, we discuss clone annotation. 
Section~\ref{sec:ReengineeringTechniques} presents the techniques for clone removal or clone based reengineering. In Section~\ref{sec:Analysis}, we  describe the analyses for the identification of potential clones as candidates for refactoring/reengineering including the visualization of clones, cost-benefit analysis and scheduling of clones for refactoring. In section \ref{sec:rootcauses}, we briefly summarize the root causes for clones followed by clone management strategies in Section \ref{sec:ManagementStrategy}.
In Section~\ref{sec:DesignSpace}, we briefly describe the design space for a clone management system. 
Our view on the challenges for industrial adoption of clone management
is presented in Section~\ref{sec:Adoption}. 
Finally, Section~\ref{sec:Conclusion} concludes the paper with a rough summary of the state of the art along with future research directions.

\section{A Systematic Review of Clone Literature}\label{sec:LiteratureReview}
There has been more than a decade of research in the field of software clones. To understand the growth and trends in the different dimensions of clone research, we carried out a quantitative review on related publications. Robert Tiras has been maintaining a repository~\cite{CloneLiterature} of scholarly articles that make significant contributions in the area. Until today, the corpus consists of 353 scholarly articles published between 1994 and 2013 in different refereed venues including Ph.D., M.Sc., and Diploma theses. The repository organizes the publications by categorizing them based on their contributions in four major sub-areas of clone research. The categories are as follows:

\begin{description}
\item [Detection] Publications in this category address techniques and tools for the detection of software clones.
	\item [Analysis] This category contains publications that perform analysis on the various traits of software clones, their etiology, existence, effects in software systems, as well as investigation of clone reengineering opportunities and implications. A majority of such publications report findings from qualitative or quantitative empirical studies. 
	\item [Management] Publications in this category address the issues, techniques and tools for the management of code clones beyond detection. 	
	\item [Tool Evaluation] This category comprises the publications that contribute to the quantitative or qualitative evaluation of the techniques and tools for clone detection.

\end{description}

\begin{figure*}[htbp]
\begin{center}
	\includegraphics[scale=0.5]{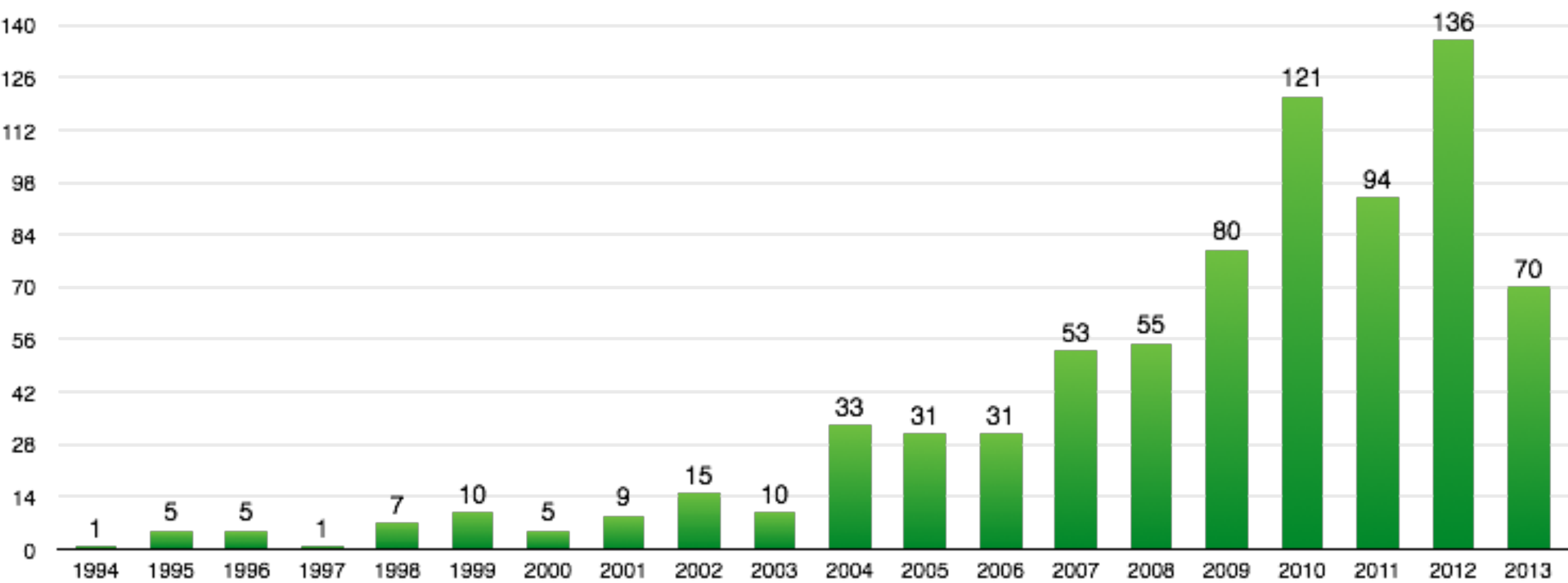}
\caption{Yearly number of distinct authors contributing to clone research}
\vspace{-0.45cm}
\label{fig:AuthorYear}
\end{center}
\end{figure*}

\begin{figure*}[htbp]
\begin{center}
\includegraphics[scale=0.4]{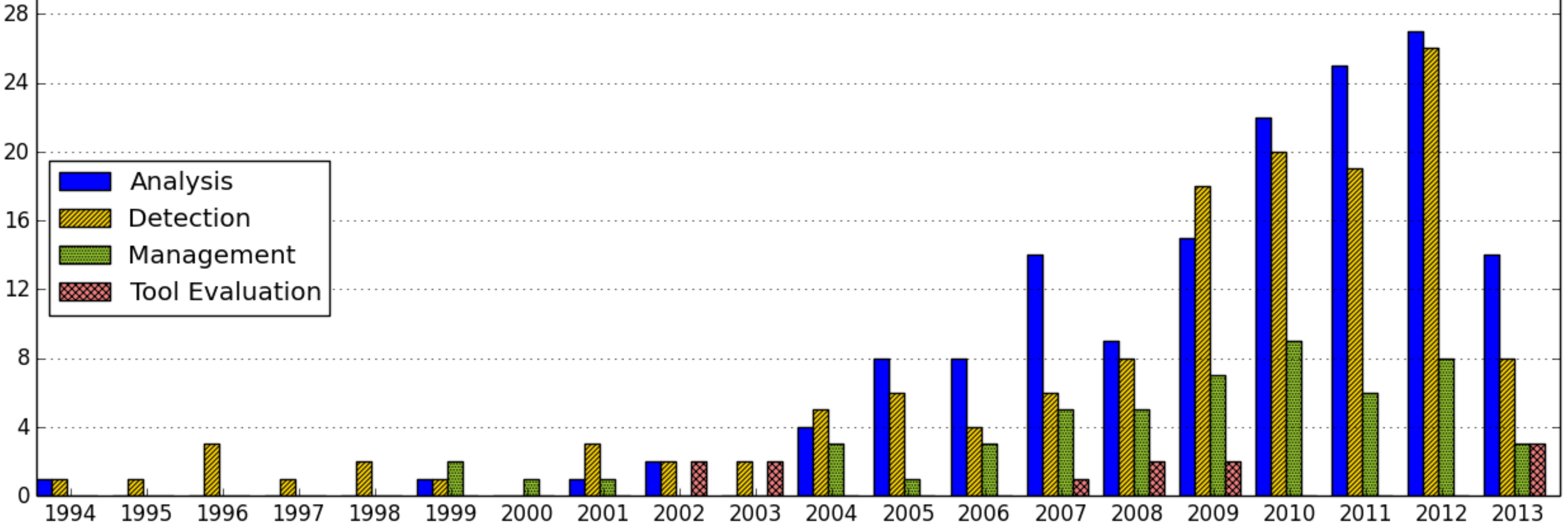}
\caption{Categories of publications on software clone research in different years}
\label{fig:CategoryYear}
\end{center}
\end{figure*}

\begin{figure}[htbp]
\begin{center}
\includegraphics[scale=0.29]{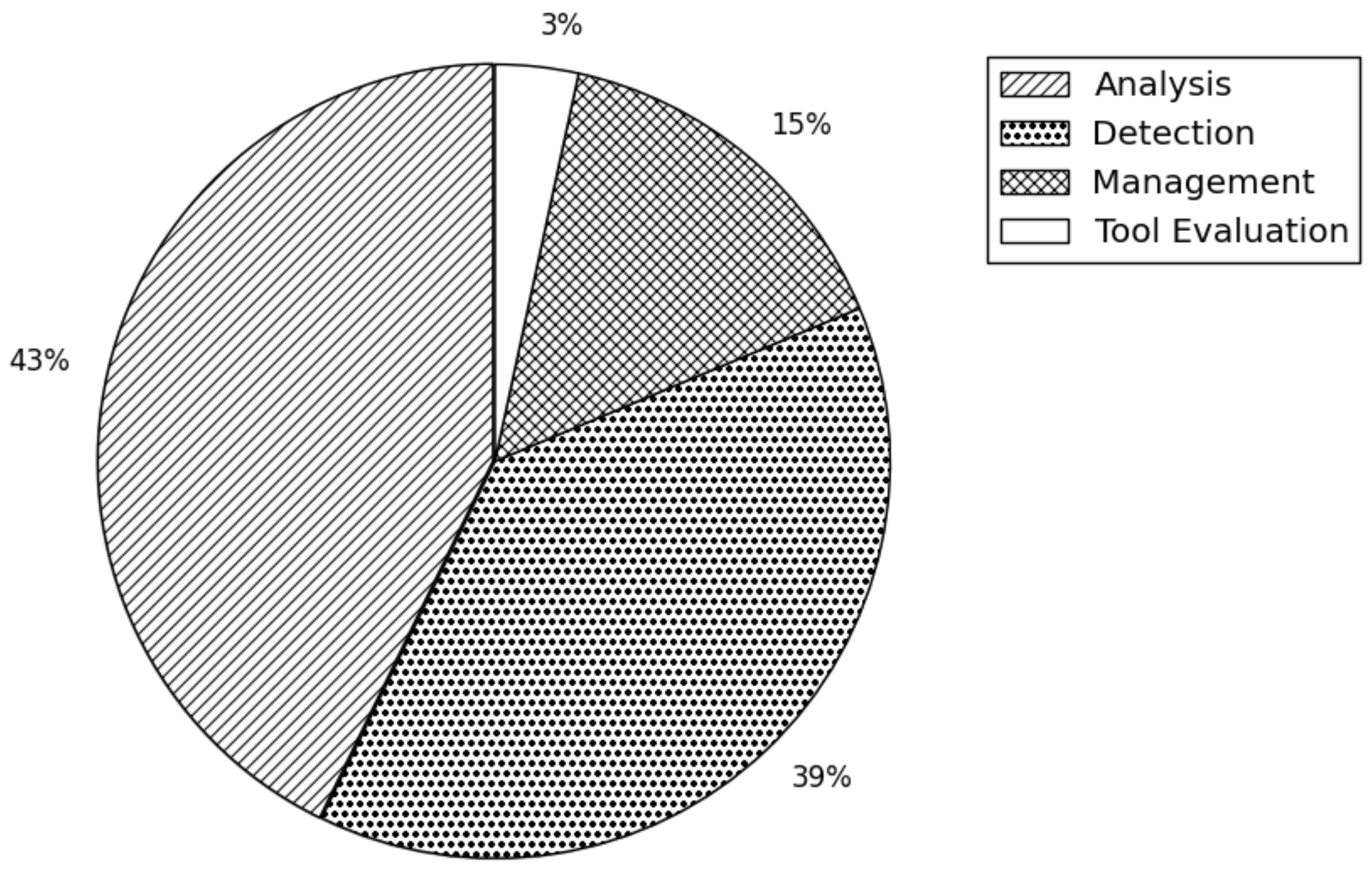}
\caption{Proportion of publications in each category over the period 1994--2013}
\label{fig:CategoryPercent}
\end{center}
\end{figure}

Figure~\ref{fig:AuthorYear} plots the number of distinct authors contributing to clone research in the years from 1994 through 2013. As the figure indicates, the clone research community has experienced a significant growth over the recent years.
In Figure~\ref{fig:CategoryYear}, we present the number of publications appear every year contributing to each of the four sub-areas of clone research. As seen in the figure, early work on software clone research was dominated by the research on clone detection with some work on analysis. In the recent years, the work on clone analysis and detection has grown significantly while clone management has emerged and growing as a significant research topic. Despite the fast growth of the clone research community, the work purely on clone management received relatively less attention compared to analysis and detection, which can be more clearly perceived from Figure~\ref{fig:CategoryPercent}. This, in combination with the realized importance of research in clone management, points to the further need and potential for research in this sub-area.

It can also be noticed from both Figure~\ref{fig:CategoryYear} and Figure~\ref{fig:CategoryPercent} that over the entire span (1994--2013) of software clone research a very few studies focused on the evaluation of clone detection techniques or tools, although more than 40 different clone detection tools have been produced realizing a wide variety of techniques~\cite{Roy_2009_ToolEvaluation}. Indeed, the detection of clones is a fundamental topic for software clone research, and the effectiveness of clone management largely depends on clone detection. 


\section{Clone Management}\label{sec:Management}
``Clone management summarizes all process activities which are targeted at detecting, avoiding or removing clones"~\cite{Giesecke_2007_CloneModel}.
Thus, clone management encompasses a wide range of categories of activities including clone detection, tracking of clone evolution, and refactoring of code clones. As support for these operations, the documentation and analysis of code clones can be regarded as parts of clone management. Moreover, clone visualization may also be an effective aid to clone analysis, and thus to clone management. 

\subsection{Definition of Code Clone}\label{sec:Clone}
Though duplicate or similar code fragments are roughly known to be code clones, the definition of clone has remained more or less vague over the last decade. The vagueness is reflected in the definition given by Ira Baxter, ``Clones are segments of code that are similar according to some definition of similarity"~\cite{Baxter_1998_CloneDr}. Despite ongoing debates in the research community, there is no consensus on a precise definition yet.
Currently, a researcher's definition of similarity is typically constrained by
the program representation and detection mechanism of his or her particular
clone detector and, hence, varies from tool to tool and also from parameter
settings controlling a tool. The least common denominator widely accepted
today is the following taxonomy, which was created in the context of a study
on comparing clone detectors~\cite{Bellon_ToolComparison}:

\begin{description}
\item [Type-1 Clone] Identical code fragments except for variations in white-spaces and comments are \emph{Type-1} clones.
\item [Type-2 Clone] Structurally/syntactically identical fragments except for variations in the names of identifiers, literals, types, layout and comments are called \emph{Type-2} clones.
\item [Type-3 Clone] Code fragments that exhibit similarity as of \emph{Type-2} clones and also allow further differences such as additions, deletions or modifications of statements are known as \emph{Type-3} clones.
\item[Type-4 Clone] Code fragments that exhibit identical functional behaviour but are implemented through very different syntactic structures are known as \emph{Type-4} clones.
\end{description}

Type-2 and type-3 clones are often collectively called \textit{near-miss clones}. There have been alternative, more elaborated taxonomies proposed by
Mayrand et al.~\cite{Mayrand_1996_Taxonomy}, Balazinska et
al.~\cite{Balazinska_1999_CloneReengineer}, and
Kontogiannis~\cite{Kontogiannis_1997_Taxonomy}, but they are not as
widely used as the simple categorization by Bellon et al.~\cite{Bellon_ToolComparison}.

A common definition is needed when empirical results are to be
compared, for instance, on effects of clones or on accuracy of clone
detectors. The difficulty to reach a consensus on a suitable
definition, however, inevitably depends also on the purpose of the
clone detection. A definition of similarity will include the ``value''
of a clone for the given task (e.g., bug fixing or refactoring). We do
not foresee the advent of a unified definition, we rather expect that
task-specific taxonomies of code similarity will emerge in the future
and studies will further differentiate contexts and purposes of
clones.

Ongoing research also attempts to deal with clones in software artifacts other than the source code~\cite{Juergens_2011_BeyondCode}, such as clones in higher level code structure~\cite{Basit_2005_HigherLevelClone},  clones in the models of formal model based development~\cite{Deissenboeck_2008_ModelClone}, in UML domain models~\cite{Storrle_2010_CloneUML}, UML sequence diagrams~\cite{Liu_2006_SDClone}, in the graph based Matlab/Simulink models~\cite{Pham_2009_ModelClone}, duplication in requirement specification documents~\cite{Juergens_2011_BeyondCode,Juergens_2010_RequirementClone}, predicting clones among domain entities \cite{RahmanARP04Domain}, and even in Spreadsheets \cite{HermansSPD13}. Definitions of clones must capture clones in all types of 
artifacts, not just source code. However, this paper focuses on the 
management of clones in the source code only.

\subsection{Clone Management Activities}\label{sec:ManagementActivity}
To manage clones, first they have to be identified. The result of clone detection forms clone documentation that records the location of code segments and their clone relationship.  If the code base changes due to ongoing development, the changes and locations of the clones need to be tracked, and the documentation needs to be updated accordingly. The clone documentation may be analyzed to determine justification of clones or to find potential clones for removal. Visualization techniques can aid such analysis. Clones that are found to have justified reason to exist may be further documented and/or annotated. The candidates for refactoring can be scheduled for modification and/or removal. Upon the application of refactoring operations, a follow up verification may examine if the refactoring caused any change in program behaviour, and in accordance, may initiate roll-back and re-refactoring. Upon completion of refactoring the clone documentation needs to be updated for consistency.

\begin{figure*}[htbp]
\begin{center}
	\includegraphics[scale=0.49]{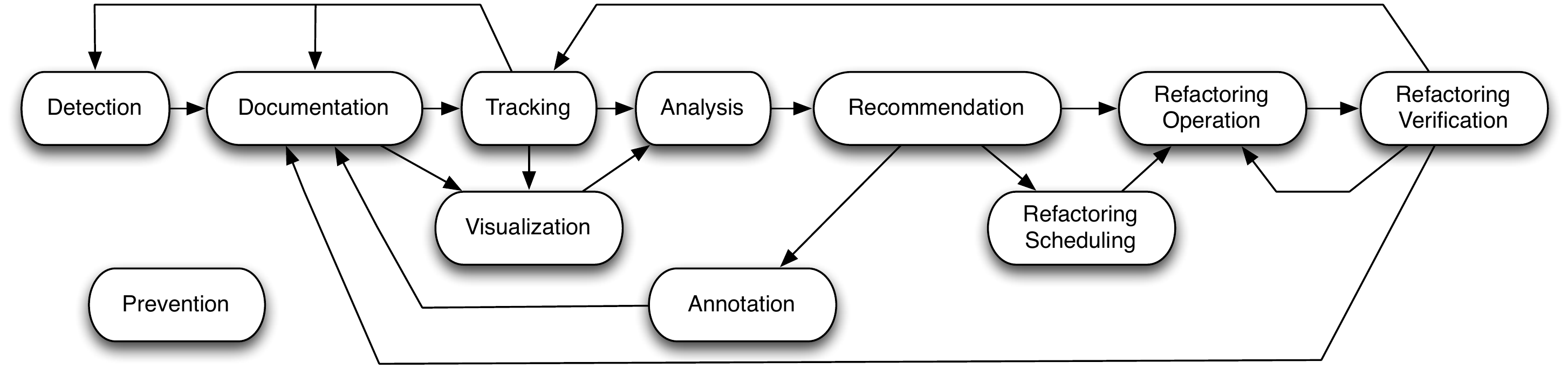}
\caption{Clone management workflow}
\label{fig:workflow}
\end{center}
\vspace{-0.15cm}
\end{figure*}

The workflow for a typical clone management system may compose all these activities according to as summarized in Figure~\ref{fig:workflow}. In the following sections, we describe the state of the art in support for each of the clone management activities.

\section{Clone Detection}\label{sec:Detection}
Over more than a decade of code clone research a number of techniques have been devised for the detection of code clones and many clone detection tools have been developed. 
%
 %
 In this section, we provide a brief summary of different clone detection techniques. More detailed descriptions of those techniques can be found in the corresponding papers and elsewhere~\cite{Roy_2007_Survey,Roy_2009_ToolEvaluation}.

\textbf{Tracking Clipboard Operations:} This technique of clone detection is based on the assumption that programmers' copy-paste activities are the primary reason for the creation of code clones. So, the technique \cite{cloneboard, Hou_2009_CnP,CPC} simply tracks clipboard activities in the editor (inside IDEs such as Eclipse) when a programmer copies a code segment and reuses by pasting it. The copied and the pasted code segments are recorded as clone-pairs.

\textbf{Metrics Comparison:}
Metrics based techniques  \cite{Lague_1997_FunctionClone,Mayrand_1996_Taxonomy} are usually used to detect function clones. The techniques are based on the assumption that similar code fragments should yield very similar values for different software metrics (e.g., cyclomatic complexity, fan-in, fan-out). Typically, for the code segments a set of metrics are gathered into vectors. The differences in the vectors are calculated, where close vectors (e.g., measured by Euclidean distance) indicate that their corresponding code fragments are clones.  

\textbf{Texual Comparison:} Text based techniques \cite{Ducasse_1999_Duploc, Johnson_1994_CloneDetect} compare program text, typically line by line, with or without normalizing the text by renaming the identifiers, filtering out the comments and differences in the layout.

\textbf{Token Based Comparison:}
In token based techniques \cite{Baker_1995_NearDuplicate,  Falke_2008_SuffixClone}, the entire program is transformed into a stream of tokens (i.e., individual units/words of meaning) through lexical analysis. Then the token stream is scanned to find similar token subsequences, and the original code portions corresponding to those subsequences are reported as clones.

\textbf{Syntax Comparison:}
Syntax comparison based techniques \cite{Baxter_1998_CloneDr, DECKARD,JSync} are developed on the fact that similar code segments should also have similar syntactic structure. Thus, the program is parsed to produce a syntax tree, where similar subtrees indicate that their corresponding code segments are clones.

\textbf{PDG Based Comparison:}
For a given program, a set of PDGs (Program Dependency Graphs) are produced based on the data and control dependencies among the statements of the program. The code segments corresponding to the isomorphic subgraphs are identified and reported as clones \cite{Higo_2009_Scorpio,Higo_2011_IncrementalPDG,Komondoor_2001_DuplicateSlicing}.

\textbf{Hash Based Comparison:}
Recently, hash based techniques are getting attention for fast and scalable detection of near-miss clones where hash values are generated from source code and processed further for finding clones \cite{Uddin_2011_SimCad, SchwarzLR12}.

\textbf{Comparison of Low Level Form of Code:} 
Instead of analyzing and comparing textual source code, the techniques analyze the lower level code (e.g., assembly code, Java Bytecode or .Net intermediate language) as obtained from the transformation by the compiler  \cite{Davis_2010_AssemblerClone, Keivanloo2013, Al-OmariKRR12}.

\textbf{Other Techniques:}
Besides the aforementioned prominent techniques for clone detection, other techniques, such as formal methods~\cite{Santone_2011_BytecodeClone}, and combination of distinct techniques~\cite{NiCad_2008} were also approached. Tracing of abstract memory states during the execution of the program was also attempted to detect semantic clones \cite{Kim_2011_MeCC}.

As listed above there have been a great many state of the art clone detectors available. However, still little is known about the usefulness of the clones detected by different clone detectors. Furthermore, evaluation of the clone detectors is still an open challenge \cite{Roy_2007_Survey,Roy_2009_ToolEvaluation} as we do not have reliable benchmarks except the tool comparison experiment of Bellon et al. \cite{Bellon_ToolComparison} and the mutation based framework of Roy,  Cordy and Svajlenko \cite{Roy_2009_Mutation, SvajlenkoRC13}. The parameter settings of the clone detectors is another threat as shown by Wang et al. \cite{Wang2013FSE} as  confounding configuration choice problem and conducted an extensive study considering six clone detectors to ameliorate the effects of the problem. Not to mention the issue of big data clone detection is a growing challenge for clone management and for many other related applications \cite{SvajlenkoKR13}.

\section{Integrated Clone Detection}\label{sec:IntegratedDetect}
There are many clone detection tools out there, each has its own
strengths and weaknesses. However, for proactive clone management, the
support for clone detection should be integrated with the development
process. Therefore, we focus on those tools that integrate clone detection
with an IDE or a version control system.

Juergens et al. developed \texttt{CloneDetective}~\cite{CloneDetective}, an open source framework to facilitate implementation of customized clone detectors. 
The framework itself is built on the infrastructure of \texttt{ConQAT}\footnote{\url{http://www.conqat.org/}}, an integrated toolkit for software quality assessment. Currently, \texttt{CloneDetective} is an integral part of \texttt{ConQAT}, which applies a suffix-tree-based technique to detect \emph{Type-1}, \emph{Type-2}, and \emph{Type-3} clones. However, beyond the detection of clones and visualization of the clone detection result, they offer no further support for clone management. 

\texttt{SimScan}\footnote{\url{http://blue-edge.bg/download.html}}, which is a parser-based tool available as plugin to Eclipse, IDEA, or JBuilder, can detect \emph{Type-1}, \emph{Type-2}, and possibly a subset of \emph{Type-3} clones. The potential of \texttt{SimScan} is also limited up to the detection code clones, not beyond that.

Giesecke~\cite{Giesecke_2007_CloneModel} proposed a generic model for describing clones. The model allowed separation of concerns among the detection, description, and management of code clones. The objective was to ease the implementation of tools to support such activities. Based on the proposed model, they implemented \texttt{DupMan}, a framework~\cite{DupMan} integrated with the Eclipse platform, and developed a prototype tool having \texttt{SimScan} as the back-end clone detector. The model and implementation is limited to the detection of clones and the representation of the clone information for persistence. 


%

\texttt{CloneBoard}~\cite{cloneboard} and \texttt{CPC}~\cite{CPC} are Eclipse plugins similar to \texttt{CloneScape} that can detect and track clones based on clip-board (copy-paste) activities of programmers. Both \texttt{CPC} and \texttt{CloneBoard}  support linked editing of clone pairs as described by Toomim et al.~\cite{Toomim_2004_CodeLink}. However, \texttt{CPC} was implemented as a framework to serve as a platform for future clone management technology, whereas, the focus of \texttt{CloneScape} was more on clone visualization and navigation, though their implementation remained incomplete. Hou et al. are developing a toolkit named \texttt{CnP}~\cite{Hou_2009_CnP} for clone management, which also detects clones based on programmers' copy-paste activities. Indeed, the current implementation of \texttt{CnP} offers very limited support for clone management, which we address in Section~\ref{sec:ConsistentRenaming}.

\texttt{SHINOBI}~\cite{SHINOBI} is an add-on to the Microsoft Visual
Studio 2005. For clone detection, it parses the source code, extracts
sequences of pre-processed tokens and creates an index using
suffix-tree based technique. \texttt{SHINOBI} internally uses \texttt{CCFinderX}'s preprocessor, and thus it can detect \emph{Type-1} and \emph{Type-2} clones only, but not \emph{Type-3}~\cite{Zibran_2012_CloneSearch}. It was developed as a client(IDE)-server(CVS) application to mainly relocate the clone detection overhead from the client to a central server. It simply displays clones of a code fragment underneath the mouse cursor, no further support for clone management is offered. \texttt{CodeRush}\footnote{\url{http://devexpress.com/Products/Visual_Studio_Add-in/Coding_Assistance/}} is a commercial add-in to the Microsoft Visual Studio for providing assistance in coding and refactoring. \texttt{CodeRush} recently introduced a new module \texttt{DDC} for the detection and consolidation of duplicated code.

 
Bahtiyar developed \texttt{JClone}~\cite{Bahtiyar_2010_MScThesis} as a plugin for Eclipse for detecting code clones from Java projects. \texttt{JClone} applies an AST based technique to detect \emph{Type-1} and \emph{Type-2} clones only.	It enables the user to  trigger the detection of clones from one or more selected files or directories. It also offers a few visualizations (i.e., TreeMap and CloneGraph views) for aiding clone analysis to some extent, but no further support for clone management beyond the detection and visualization of clones. 
	
Nguyen et al. developed \texttt{JSync}~\cite{JSync} as a plugin to the SVN version control system. Earlier prototypes of \texttt{JSync} appeared as \texttt{Clever}~\cite{Clever} and \texttt{Cleman}~\cite{Nguyen_2008_Cleman}. \texttt{JSync} detects clones based on similarities among the feature vectors computed over AST representation of the code fragments. 
\texttt{JSync} incorporates some useful features for clone management, which are discussed in Sections~\ref{sec:Tracking} and \ref{sec:ReengineeringTechniques}.

\texttt{CPD}\footnote{\url{http://pmd.sourceforge.net/cpd.html}} is a part of the Java source code analyzer, \texttt{PMD}. 
\texttt{SDD}~\cite{SDD} is a clone detection algorithm based on n-neighbour distance, index and inverted index. An implementation of \texttt{SDD}\footnote{\url{http://wiki.eclipse.org/index.php/Duplicated_code_detection_tool_(SDD)}} is also freely available as a plugin to Eclipse.
 \texttt{Simian}\footnote{\url{http://www.harukizaemon.com/simian}} is another clone detector available as a plugin to Eclipse.
Another Eclipse plugin, \texttt{CloneDigger}\footnote{\url{http://clonedigger.sourceforge.net/download.html}}, applies an approach based on AST, suffix tree, and anti-unifcation for detecting clones in source code written in Java or Python.  
Tairas and Gray~\cite{Tairas_AST_suffix} also developed a suffix-tree based clone detector as a plugin for the Microsoft Phoenix framework. 
Despite the integration with IDEs all these tools offer no support for clone management except for the detection of only \emph{Type-1} and \emph{Type-2} clones~\cite{Bettenburg_2010_CloneChange}.

Another Eclipse plugin, \texttt{CloneDR}\footnote{\url{http://www.semdesigns.com/Products/Clone/}},  is an AST-based clone detector that can detect \emph{Type-1} and \emph{Type-2} clones. Besides clone detection, \texttt{CloneDR} offers support for clone removal as further discussed in Section~\ref{sec:RefactoringPatterns}.
\texttt{CeDAR}~\cite{CeDAR} can incorporate the results from different clone detection tools (e.g., \texttt{CCFinder}, \texttt{CloneDR}, \texttt{DECKARD}, \texttt{Simian}, or \texttt{SimScan}) and can display properties of the clones in an IDE. \texttt{CeDAR} offers no further support for clone management, except that those clone properties may be useful for clone analysis. Moreover, it may suffer from the limitations of the underlying clone detector used internally. 
Recently, Zibran and Roy~\cite{Zibran_2012_CloneSearch} developed an Eclipse plugin to facilitate focused search for clones of a selected code fragment. They applied a suffix-tree-based k-difference hybrid approach to detect both exact (\emph{Type-1}) and near-miss (\emph{Type-2} and \emph{Type-3}) clones. They are also extending their tool towards a versatile clone management tool~\cite{Zibran_2011_CloneManage}.

While we see that there are a number of IDE-based clone detection tools available, there are only a few that in fact can deal with \emph{Type-3} clones. Furthermore, as we will see in the following sections that there is still a marked lack for different clone management features in these IDE-based tools. Researchers possibly should first conduct user studies of what sort of features are needed for effective clone management and then start building tools that would help developers and maintenance engineers in dealing with different types of clones.

\section{Clone Documentation}\label{sec:Documentation}


Different clone detectors report the results of clone detectors in different formats such as XML, HTML, and plain text.  There are variations in the reported information as well. Some clone detectors report clone pairs only, while some other tools report clones in terms of clone groups. Such variations make it difficult for data exchange between clone detectors, which also adds to the challenges in head-to-head empirical comparison of clone detectors. To minimize the differences in the presentation of clone information, Harder and G\"ode~\cite{Harder_2011_CyCloneRCF} recently proposed the \emph{Rich Clone Format (RCF)}, an extensible schema based data format for storage, persistence, and exchange of clone data.


Duala-Ekoko and Robillard~\cite{Ekwa_2010_CRD} proposed clone region descriptor (CRD) to describe clone regions within methods in a way that is independent of the exact text of the clone region or its absolute location in a file. 
However, such a scheme has a number of limitations. First, small changes in the code corresponding to the \textit{<anchor>} (e.g., termination condition of loop, branching predicate of conditional statements) will invalidate the CRD. Second, the scheme is vulnerable to nesting levels, and thus a simple addition or removal of nesting level will invalidate the CRD. Third, the association of `else' blocks with the closest `if' block prevents the CRD scheme differentiating between the two types of blocks. 
Most importantly, the use of the CRD scheme did not save \texttt{CloneTracker}~\cite{Ekwa_2010_CRD} from re-invoking the underlying clone detector to identify possible changes in the clones, though the computational expense of re-detection was indicated as one of the motivations behind the design of CRD.

The above discussion indicates that the line and column information, or the abstract level CRD based documentation of clone regions are more or less vulnerable to changes in the evolving code.  To overcome such sensitivity to code change, marker based tagging support in IDEs like Eclipse can be used for clone documentation. Such tagging of clones can provide built-in support for accommodating changes in the source files~\cite{Chiu_2007_BeyondDetection}. Further investigation may be required to verify this possibility.

%
%
Capturing the location of clones reliably is necessary for tool
comparisons and also for tracing clones over subsequent versions.  If
tools are to be integrated from different vendors, an agreed way to
document clones is required. RCF is a step towards a common
format \cite{Harder_2011_CyCloneRCF}, but it does not address all needs
\cite{Kapser:IWSC:2012}. A common conceptual model for clone
information is a major challenge because of competing requirements
(e.g., it should be both generic and efficient)
\cite{Harder:IWSC:13}. There has been some progress towards a unified
model \cite{Kapser:IWSC:2012}. We expect real practical progress to
happen, however, only if different research teams actually start to
exchange data -- and not just between two teams but among many
teams. We do not see this happening at the moment except for
exchanging benchmark data for clone detectors and for that use case,
RCF seems to be sufficient.

\section{Clone Tracking}\label{sec:Tracking}
During the development of an evolving software system frequent changes take place in the code base. Such changes may introduce new code segments that might form new clones. Moreover, changes in source files may invalidate the clone regions necessitating corresponding updates in the recording of clone information. Such updates can be accomplished in two ways: Re-detection and incremental detection. 

\textbf{Re-detection:} The detection of clones from the entire system may be invoked every time the code changes. This approach may incur too much overhead as the detection of code clones in a fairly large system can be computationally expensive. Hence, the approach might not be suitable for proactive clone management.

\textbf{Incremental Detection:}
A better approach can be incremental detection, where only the source code in the modified portion of the code base is examined for any clones and the outcome is accumulated with the previously preserved clone detection results.
%
%
Not many attempts were made towards incremental clone detection.
The first attempt was made by G\"ode and Koschke~\cite{Gode_2010_IncrementalEvolution}. They proposed a suffix tree based detector \texttt{iClone} for incremental detection of clones in subsequent versions of a given system. It detects \emph{Type-1},  \emph{Type-2}, and \emph{Type-3} clones.

Hummel et al.~\cite{Hummel_2010_IncerementalIndex} proposed an index-based incremental clone detection approach for \textit{Type-1} and \textit{Type-2} clones. Higo et al.~\cite{Higo_2011_IncrementalPDG} proposed an incremental one based on PDGs, where PDGs are generated from the analysis of control and data dependencies in the program code targeting even semantic clones. However, PDG based techniques are computationally expensive and they often report \emph{non-contiguous} clones that may not be perceived as clones by a human evaluator \cite{Bellon_ToolComparison}.
%
%
%
%
%
The clone tracking approach of
\texttt{JSync}~\cite{JSync,Nguyen_2009_IncrementalVector} appears to
be computationally elegant. \texttt{JSync} preserves the clone-groups
and $N$ buckets obtained from the initial clone
detection.  Since \texttt{JSync} is implemented as a plugin to SVN, the change information of the source files are readily available, and based on that information \texttt{JSync} can determine the fragments modified, added to, or deleted from the code repository. \texttt{JSync} then removes from the clone-groups those fragments that were changed or deleted. Then the LSH technique is applied to the newly added and modified code fragments to place them in the buckets. Then the fragments in each bucket are compared pair-wise to update the clone-groups. Thus, the clone detection technique of \texttt{JSync} appears to be inherently incremental and consequently computationally efficient for tracking clones.
We envision that other classes of techniques (cf., Section~\ref{sec:Detection}) will also
have incremental variants as there is a clear need for scalable, fast
and near-miss incremental detection techniques for efficient clone
management.

\section{Analysis of Clone Evolution}\label{sec:Evolution}
Software development and maintenance in practice follow a dynamic process. With the growth of the program source, code clones also experience evolution from version to version. Many studies have been conducted to date for understanding the overall evolution~\cite{Gode_2010_IncrementalEvolution, Zibran_2011_CloneForecast, SahaRSP13Type3Evolution, LillianeLatePropagation}, stability of cloned code~\cite{Bakota_07_CloneSmell, HarderG13StableCode, Hotta_2010_CloneStable, Krinke_2008_CloneStability, Mondal_2012_CloneStability, Mondal2012Stability, Mondal2013InsightInto}, the relation of clone evolution with software faults~\cite{LillianeLatePropagation, XieKZ13Fault, XieWCRE13Fault, Gode_2011_RiskClone}, and other characteristics of clone evolution.  While such studies inform the characteristic and impact of code cloning, further in-depth analyses that investigate the change patterns in the evolution of individual clone fragments can suggest techniques for optimizing clone management including refactoring and removal. Because there is already a recent survey on clone evolution~\cite{Pate_2011_EvolutionSurvey}, we keep this section brief with specific focus on the evolution of individual clone fragments and their change patterns.

Kim et al.~\cite{Kim_2005_Empirical} first coined the term \emph{``clone genealogy"}, which refers to a set of one of more lineage(s) originating from the same clone-group. A \emph{clone lineage} is a sequence of clone-groups evolving over a series of versions of the software system. 
%
%
%
To map clones across subsequent versions of a program (i.e., extraction of clone genealogies) several approaches have been proposed in the literature. 
While most of these approaches \cite{LillianeLatePropagation,Bakota_07_CloneSmell,Kim_2005_Empirical,Saha_2010_Genealogy} focused on genealogies of \emph{Type-1} and \emph{Type-2} clone genealogies, \texttt{gCad}~\cite{Saha_2011_gCad} is the only \emph{Type-3} clone genealogy extractor to date released as a separate tool \cite{SahaRS13gCadTool}.

Studies~\cite{Bakota_07_CloneSmell,Bettenburg_2010_CloneChange} on clone change patterns revealed that inconsistent changes in clones sometimes caused program faults. Moreover, late propagation is reported to have even more significant correlation with software defects and thus concluded to be ``more risky than other clone genealogies"~\cite{LillianeLatePropagation}.
On the basis of clone genealogies, a number of studies~\cite{Kim_2005_Empirical, Saha_2011_gCad, Saha_2010_Genealogy, SahaRSP13Type3Evolution, ZibranSRS13CloneRemoval} have been conducted to explore the change patterns and characteristics in the evolution of individual clone fragments. Some of the findings from those studies compliment one another, while some of the results derived from those studies appear to be contradictory. Therefore, more studies in larger scale are still necessary to confirm the agreeing observations and to shed light on the contradictory findings. 
%

Studies on clone change patterns using a genealogy
model can suffer from a number of issues.
%
First, due to the threshold based similarity measure used in practice for \emph{Type-3} clones, there remains an open question on the appropriate value for the threshold. Moreover, for \emph{Type-3} clones, is it an appropriate practice to group \emph{Type-3} clones into different disjoint classes? If not, the traditional notion of genealogy cannot apply to \emph{Type-3} clones. Can we devise a more appropriate alternative? 
Second, a genealogy can be characterized as inconsistently changed if only a single clone over the entire length of the genealogy experiences even for very minor inconsistent changes. 
To draw a better picture, we may capture information such as, what portion of clones in clone groups change in how many versions, and how large the changes are. 
%
Finally, from the correlation between late propagation and software defects, can we really derive a causal relationship concluding that late propagation is riskier than other clone genealogies? Inconsistent changes are believed to often cause defects, and clones may disappear from a genealogy due to inconsistent changes.  Later modifications, which could be even bug fixing activities, may cause changes in the disappeared clone to sync it to its original clone-group. In such a scenario, late propagation actually contributes in repairing the defect introduced from inconsistent changes. Thus, we believe, late propagation can really play a dual role, and more studies are necessary to distinguish them.


Visualization support can aid analysis of clone evolution, and thus different techniques and tools have been proposed for visualizing properties of clone evolution including the genealogy model. 
Adar and Kim~\cite{Ader_2007_SoftGUESS} developed \texttt{SoftGuess}, a system for clone evolution exploration that supports three different views. 
The genealogy browser offers a simple visualization of clone evolution where nodes represent clones, arranged from left to right, and those that belong to the same class are arranged vertically in the same position. Thus, each column represents a version. A link between a pair of node reflects the predecessor and successor relationship during the evolution of the software. The encapsulation browser shows how clones within a clone group are distributed in different parts of a system and how they fit in the hierarchical organization of the software system by visualizing the containment relationship through a tree structure. Finally, the dependency graph describes how the nodes (package, class or method) within a version are evolved from other nodes and how they evolve in the next version. In addition, \texttt{SoftGUESS} also supports charting and filtering mechanisms based on Gython, an SQL-like query language. However, \texttt{SoftGUESS} lacks an `overview' feature and requires user interaction for data reduction through queries. Although a query is a powerful mechanism to identify important patterns of cloning, formulating queries could be difficult as this requires more cognitive effort from the developers.

Harder and G\"ode~\cite{Harder_2011_CyCloneRCF} developed a multi-perspective tool for clone evolution analysis, called \texttt{CYCLONE}. It offers five different views to analyze clone data stored in an RCF file, where RCF is a binary format to encode clone data including the evolutionary characteristics. The evolution view in \texttt{CYCLONE} visualizes clone genealogies using simple rectangles and circles to denote software entities. Each circle represents a clone fragment arranged in a set of rows where each row represents a particular version of the software. The clones that belong to the same clone class are packed within a rectangle. Finally, lines represent the evolution of a clone fragment. In addition, the view employs colors to distinguish types and the changes of the clones. Although the view highlights many important evolutionary characteristics, the volume of data produced by the genealogy extractor still limits its usefulness, thus calls for overview and filtering mechanisms. A similar visualization support is available in \texttt{VisCad}~\cite{AsaduzzamanRS11IWSC}, with additional flexibility of metric-based filtering of genealogies.

While there have been a good number of studies (c.f., Section \ref{sec:Analysis}) on visualization of clones in a single version of a software system, we still need further studies to figure out useful techniques for visualizing clone evolution from management perspective. For example, what sort of visualizations are useful for clone management activities and what are their implications in the context of real world software development? We need to understand the claims and believes about code clones \cite{ChatterjiCK12ClaimsBeliefts} including empirical evidence from developers' perspective \cite{ChatterjiCKH13Replicated, ZibranSRS13CloneRemoval} and then need to design the visualization techniques appropriately.  It is also important to understand developers' intent when designing such tools \cite{ChatterjiCK13Intent}. 

Recently, Saha et al.~\cite{Saha_2011_VisualizeCloneEvolution} presented an idea for clone evolution visualization using the popular scatter plot. In their proposed approach, scatter plots show the clone pairs associated within a pair of software unit (file, directory or package). Based on the type of clone genealogies they are associated with, clone pairs are rendered with different colours. Selecting a clone pair through user interactions (double clicking on a clone pair in the scatter plot) shows the associated genealogy in a genealogy browser. The proposal facilitates developers or maintenance engineers to identify evolutionary change patterns of the clone classes in a particular version and then provide a way to call for genealogy browser to dig deeper. However, it does not provide overall characteristics of the genealogies. Moreover, due to the large number of clone pairs, selection and useful pattern identification in such a scatter plot can be difficult, which is why different variants of the traditional scatter plots appeared in the literature~\cite{Cordy_2011_LiveScatterplot}.

\section{Clone Annotation}\label{sec:Annotation}
The developers often deliberately create clones, for example, to enable independent evolution of similar implementations. During the clone management process, the developer may not want to refactor/remove those clones, and may want to mark those to indicate such decisions so that they will not have to encounter those same sets of clones over and over.  Moreover, the decision needs to be documented and shared among different programmers, and there should be facilities for the developers to review those clones at a later time, in case they want to re-evaluate their management decision.
To the best of our knowledge, such a feature is found only in \texttt{JSync}~\cite{JSync}, which allows the developer to annotate pairs of clones for avoiding future encounters.

Although there are several ideas and implementations of
clone-evolution visualization, there is not enough empirical assessment
of these. We also believe that further progress can be achieved by
studying existing work in information visualization. 

 \section{Techniques for Reengineering/Refactoring of Clones} \label{sec:ReengineeringTechniques}
The investigations of opportunities for clone based reengineering and refactoring of clones for their removal have suggested techniques such as generics, design patterns, software refactoring patterns, and synchronized modifications of code clones.

\textbf{Generics and Templates:}
Basit el al.~\cite{Basit_2005_CloneGenerics} investigated the potential of generics in removing code clones. They carried out two case studies on the \emph{Java Buffer Library} and the \emph{C++ Standard Template Library (STL)}. The \emph{Java Buffer Library} was found to have 68\% redundant code, and using generics they were able to remove only 40\% of them. Though, they performed little better for the \emph{C++ STL}, they  concluded that the constraints of language constructs limit the applicability of generics in clone removal. They further hypothesized that meta level parameterizations might perform better as they are relatively lesser restrictive than generics or templates.

The hypothesis on the potential of meta level parameterizations was addressed by Jarzabek and Li~\cite{Jarzabek_2006_CloneUnify} in a later study. They also used the \emph{Java Buffer Library} for their case study. They applied a generative programming technique using \emph{XVCL (XML-based Variant Configuration Language)}\footnote{\url{http://xvcl.comp.nus.edu.sg/}} to represent similar (but not necessarily identical) classes and methods in generic and adaptable form. Using the technique they were able to eliminate 68\% of the code from the original \emph{Java Buffer Library}.

\textbf{Consistent Renaming:} \label{sec:ConsistentRenaming}
Programmers often perform modifications after copy-pasting a code fragment. Such modification typically include renaming of identifiers according to the new context of the cloned code. IDEs like Eclipse provide necessary support for consistently renaming an identifier and all its references within scope. Jablonski and Hou developed \texttt{CReN}~\cite{CReN} as a plugin to Eclipse that can check for any inconsistencies in the renaming of identifiers within a code fragment and suggest modifications for making the renaming consistent. 
%
%

Since \texttt{JSync}~\cite{JSync} is developed as a plugin to the SVN version control system, it can exploit the change information between versions of Java source files to determine whether any changes occurred in cloned code regions. 

\textbf{Refactoring Patterns}\label{sec:RefactoringPatterns}
Fowler in his book~\cite{Fowler_1999_Refactor} presented 72 patterns for refactoring source code in general for the removal of code smells. Over time, the number of refactoring patterns has increased to 93, and a \emph{refactoring catalog}\footnote{Catalog of OO refactoring patterns: \url{http://refactoring.com/catalog/}} is maintained that lists and describes them all. Among those general software refactoring patterns~\cite{Fowler_1999_Refactor}, \textit{Extract method}, \textit{Move method}, \textit{Pull-up method},  \textit{Extract superclass}, \textit{Extract utility-class}, and \textit{Rename refactor} patterns are found to be suitable for clone refactoring, as suggested by earlier research~\cite{Higo_2004_RefactoringClone,Lee_2010_RefactorSchedule,Schulze_2009_CloneRemoval,Yoshida_2005_CloneRefactor,Zibran_2011_RefactorSchedule, ZibranR13RefscheduleJournal}. 
Details about these refactoring patterns can be found in the \emph{refactoring catalog} and elsewhere~\cite{Fowler_1999_Refactor}.

Besides these prominent refactoring patterns, other low level refactoring operations such as {\em identifier renaming}, {\em method parameter re-ordering}, {\em changes in type declarations}, {\em splitting of loops}, {\em substitution of conditionals}, {\em loops}, {\em algorithms}, and {\em relocation of methods or fields} may be necessary to produce generalized blocks of code from near-miss clones~\cite{Zibran_2011_RefactorSchedule}. Kerievsky~\cite{Kerievsky_2004_RefactorPattern} proposed the \emph{chained constructor} refactoring pattern\footnote{Catalog of 27 refactoring patterns from J. Kerievsky's book: \url{http://industriallogic.com/xp/refactoring/catalog.html}}, to eliminate duplicated code from the constructors of the same class~\cite{Nasehi_2007_ChainedConstructor}. Other refactoring patterns that can be found in the literature are some sort of variants or compositions of the aforementioned object-oriented refactoring (OOR) patterns.
Other than the OOR patterns, Schulze et al.~\cite{Schulze_2008_CloneClassification} proposed three aspect oriented patterns described as \emph{extract feature into aspect, extract fragment into advice}, and \emph{move method from class to interface}. 

Type-3 clones remain a challenge for automated clone refactoring
because they have difference that cannot be eliminated with a simple
rename refactoring. Here, an additional step is required to abstract
from the difference in a way that enables an extract-method
refactoring. Anti-unification used to detect clones may help in
refactoring, too, in certain situation. We expect progress for some
\emph{Type-3} clones at least. It is an interesting question how far we can
get. A couple of recent studies \cite{KrishnanWCRE13Unification, BazrafshanK13CloneRemoval} also call for further studies on clone refactoring including the refactoring of task specific near-miss clones. Indeed, clone maintenance support could be increased by unifying clone detection and refactoring activities \cite{TairasG12Refactoring} and we need to focus more on such studies.

\section{Analysis and Identification of Potential Clones for Refactoring}\label{sec:Analysis}
For the purpose of finding and characterizing code clones suitable for refactoring, reengineering, or removal, in depth analysis of the various properties of the clones and their context is required. Clone visualization has been proven to be effective in aiding such analysis. Therefore, we first discuss the tools and techniques for code clone visualization, and then we present the findings from analysis of code clones in search for clone based reengineering opportunities.

\subsection{Visualization of Distribution and Properties of Clones}

Almost all the clone detection tools report clone information in the
form of clone pairs and/or clone groups in a textual format where
only the basic information about the clones such as the file name,
line number, starting position, ending position of clones are
provided. The returned clones also differ in several contexts such
as types of clones, degree of similarity, granularity and size.

Moreover, there is a huge amount of clones in large systems. For
example, {\em CCFinder} resulted 13,062 clone pairs for Apache httpd
\cite{Kapser_2009_PhDThesis}. Because of the insufficient
information on the returned clones, their various contexts, and their sheer
number, the presentation of clones becomes difficult. For the
proper use of the detected clones, especially for clone management,
the aid of suitable visualization is crucial. In the following,
we list some of the visualization approaches that have been proposed
in the literature.

A major challenge in identifying useful cloning information is to handle the large volume of textual data returned by the clone detectors. To mitigate the problem, a number of visualization techniques, filtering mechanisms and support environments are proposed in the literature. Jiang et al.~\cite{Jiang_2006_VisualizeCohesionCoupling} categorized the proposed clone presentation techniques based on two dimensions. The first dimension refers to the level at which the entities are visualized (such as at the code segment level or file level or subsystem level). The second dimension refers to the type of clone relation addressed by the presentation, that is, whether clones are shown at the clone pair level or grouped into clone classes or super clones. A \emph{super clone} is an aggregated representation of multiple clone groups between the same source entities (e.g., file). 

Johnson~\cite{Johnson_1994_VisualizeRedundancy} used the popular Hasse diagram to represent textual similarity between files. Later, he also proposed hyper-linked web pages to explore the files and clone classes~\cite{Johnson_1996_NavigateRedundancy}. Cordy et al.~\cite{Cordy_2004_DetectNearMiss} used HTML for interactive presentation of clones where overview of the clone classes is presented in a web page with hyperlinks and users can browse the details of each clone class by clicking on those links. Although such representations offer quick navigation, they cannot reveal the high level cloning relations. 

A set of polymetric views~\cite{Rieger_2004_CodeDuplicate} were also proposed in the literature that permit encoding multiple code clone metrics in visual attributes. Among various visualizations, scatter plot is quite popular and capable of visualizing inter-system and intra-system cloning~\cite{Cordy_2011_LiveScatterplot,Livieri_2007_DCCFinder}. However, the size of the scatter plot depends on the size of input rather than the amount of cloning. Thus, using a scatter plot for visualizing cloning relation of a large software system may become challenging due to the large size of the plot. 

Moreover, in scatter plot, non-contiguous sections that contain the same clone cannot be grouped together. 
To overcome this limitation, Tairas et al.~\cite{Tairas_2007_VisualizeResult} proposed a graphical view of clones (also known as Visualizer view) that represents each source file as a bar and clones within the files are represented with stripes. Clones belong to the same class are encoded with the same color.

Jiang et al.~\cite{Jiang_2006_StudyFramework} extended the idea of cohesion and coupling to code clones and proposed a visualization that uses shape and color to encode the metric values. They also developed a framework~\cite{Jiang_2006_StudyFramework} for large scale clone analysis and proposed another visualization, called a clone system hierarchical graph that shows the distribution of clones in different parts (with respect to the file-system hierarchy) of a system. Fukushima et al.~\cite{Fukushima_2009_DiffusedMetric} developed another visualization using graph drawing to identify diffused (scattered) clones. Here, nodes represent the clones. Those nodes that are located in the same file are connected with edges to form a clone set cluster. Nodes that connect different clone set clusters are called diffused clones (have cloning relation in different files implementing different functions).

\texttt{Gemini}~\cite{Gemini_2002_Ueda} is an example of a clone support environment that uses CCFinder for clone detection and can visualize clone relationships using scatter plots and metric graphs. Kapser and Godfrey developed \texttt{CLICS}~\cite{Kapser_2005_CLICS,Kapser_2006_SupportAnalysis}, another tool for clone analysis. \texttt{CLICS} can categorize clones based on their previously developed clone taxonomy~\cite{Kapser_2004_CloneCategorization} and support query based filtering. Tairas et al.~\cite{Tairas_2007_VisualizeResult} developed an Eclipse plug-in that works with \texttt{CloneDR} as a clone detector and implements the visualizer view along with general information and detected clones list views. 

\texttt{Clone Visualizer}~\cite{Zhang_2008_QueryVisualization} is an Eclipse plugin that works with \texttt{Clone Miner} to detect clones. In addition to supporting clone visualization through stacked bar charts and line graphs, it supports query based filtering. \texttt{CYCLONE}\footnote{\url{http://softwareclones.org/cyclone.php}}~\cite{Harder_2011_CyCloneRCF} is another clone visualizer that supports single and multi-version program analysis and uses RCF (Rich Clone Format)~\cite{Harder_2011_CyCloneRCF} file as an input. A separate viewer application named \texttt{RCFVIEWER}\footnote{\url{http://www.softwareclones.org/}} is also developed for the visualization of clone information stored in RCF. Recently, Xing et al. \cite{Xing2013WCREContextual} proposed \textit{CloneDifferentiator}  that identifies contextual differences of clones and allows developers to formulate queries to distill candidate clones that are useful for a given refactoring task.

As can be noted, all the visualization techniques focus on visualization of clone pairs or clone groups with respect to their dispersion in the file-system hierarchy only. However, the cost-benefit analysis of code clone refactoring (Section~\ref{sec:Scheduling}) takes into account the distribution of clones in the inheritance hierarchy. Therefore, from the perspective of clone removal or refactoring, the visualization of the clones with respect to the inheritance hierarchy can offer useful insights, and future work in clone visualization should address this possibility. 

For visualization of clone evolution, the
proposed techniques for visualizing clones of one version of a system lacks 
empirical assessment mostly. We possibly need use-case specific visualizations with empirical support as of Live Scatterplots \cite{Cordy11ScatterPlot}. We thus expect to have more empirical user studies as the
field matures. Furthermore, clone visualization from big data is also badly needed \cite{ForbesKR12Doppel}.

\subsection{Cost-benefit Analysis and Scheduling of Refactoring}\label{sec:Scheduling}
Not much research has been done towards cost-benefit analysis of code
clone refactoring and their scheduling. Bouktif et
al.~\cite{Bouktif_2006_Refactor} first proposed a simple effort model
for the refactoring of clones in procedural code. 
%
%
Zibran and Roy~\cite{Zibran_2011_RefactorSchedule_ICPC, ZibranR13RefscheduleJournal, Zibran_2011_RefactorSchedule} proposed a more comprehensive effort model for estimating clone refactoring efforts. They formulated scheduling of code clone refactoring as a constraint satisfaction optimization problem and applied constraint programming (CP) technique to compute an optimal solution of the problem. 

Lee et al.~\cite{Lee_2010_RefactorSchedule} applied ordering messy GA (OmeGA) to schedule refactoring of code clones. Mondal et al. \cite{MondalWCRE14Ranking} proposed an automatic way of ranking clones for refactoring through mining association rules of the evolving clones. Juergens and
Deissenboeck \cite{Juergens:IWSC:10} described a detailed analytic cost model based on
potential effects of clones on different maintenance activities. The existing models make several implicit and
explicit assumptions and do not give concrete values for weights
included in the formulae. 

Overall, we know too little about the real costs incurred by clones
and the risks and benefits of refactorings and other measures to
compensate the negative effects of clones for a realistic cost
model. We hardly know the factors influencing the costs. Only through
a series of empirical field studies and experiments will we ever get
closer to such a cost model. We remain skeptical as to whether we will
ever get close enough given the many variables influencing the costs
and gains of clones.

\section {Root causes for Code Duplication} \label{sec:rootcauses}
\label{secReasonsCloning} Code clones do not occur in software
systems by themselves. They are created. There are several factors that might force or
influence the developers and/or maintenance engineers in cloning
code in a system. In order to manage clones properly, we need to study
the root causes for their creation. In the following we list some of
the potential root causes.
%

\textbf{Development Strategy} Clones can be introduced in software
systems due to the different reuse and programming approaches. Reusing
code, logic, design and/or an entire system (as in product lines
\cite{Dubinsky:CSMR:13}) are the prime reasons of code
duplication. Reusing existing code by copying and pasting (with or
without minor modifications) is the simplest form of reuse mechanism
in the development process which results in code duplication. It is a
fast way of reusing reliable semantic and syntactic constructs. 

The term {\em Forking} is used by Kapser and
Godfrey \cite{Kapser_2008_CloneHarmful} to mean the reuse of similar
solutions with the hope that they will diverge significantly
with the evolution of the system. For example, when creating a
driver for a hardware family, a similar hardware family may already
have a driver, and thus can be reused with slight modifications.
Similarly, clones can be introduced when porting software to new
platforms and functionality
and logic can be reused if there is already a similar solution
available. 


%
%
%
%
\textbf{Maintenance Benefits} Clones are also introduced in the
systems to obtain several maintenance benefits. One of primary factors could be reducing risk in developing new code. Cordy
\cite{Cordy_2003_Adoption} reports that clones do frequently occur
in financial software as there are frequent updates/enhancements of
the existing system to support similar kinds of new functionalities.
Financial products do not change that much from the existing one,
especially within the same financial institutions. The developer is
often asked to reuse the existing code by copying and adapting to
the new product requirements because of the high risk (monetary
consequences of software errors can run into the millions in a
single day) of software errors in new fragments and because existing
code is already well tested (70\% of the software effort in the
financial domain is spent on testing). Introduction of new bugs can be avoided in critical system
functionality by keeping the critical piece of code untouched
\cite{GravesToddL00PredictingFault}. For keeping the software architecture clean and understandable sometimes clones are  
intentionally introduced to the system \cite{Kapser_2008_CloneHarmful}.



%


\textbf{Overcoming Underlying Limitations} Clones can be introduced due to
limitations of the programming language, especially when the language in
question does not have sufficient abstraction mechanisms such as inheritance, generic types (called templates in C++) or
parameter passing (missing from, e.g., assembly language and COBOL)
and consequently, the developers are required to repeatedly
implement these as idioms. Such repeating activities may create
possibly small and potentially frequent clones
\cite{BasitHamid05STL}. 

There are also several limitations associated with the programmers for which clones are
introduced in the system. For example, it is generally
difficult to understand a large software system. This forces the
developers to use the example-oriented programming by adapting
existing code developed already. Furthermore, often developers are
assigned short time frames in completing tasks. 
Due to such time limits, developers look for an easy way of solving the problems
at hand and consequently look for similar existing solutions and consequently clones are introduced in software. 
Sometimes the productivity of a developer is measured by the number
of lines he/she produces per hour. In such circumstances, the
developer's focus is to increase the number of lines of the system
and hence tries to reuse the same code again and again by copying
and pasting with adaptations instead of following a proper
development strategy.  
Sometimes the developer is not familiar with the problem domain at
hand and hence looks for existing solutions of similar problems. 


\textbf{Cloning By Accident} Clones may be introduced into software even by
accidents. The use of a
particular API normally needs a series of function calls and/or
other ordered sequences of commands. For example, when creating a
button using the Java SWING API, a series of commands is to create
the button, add it to a container, and assign the action listeners.
Similar orderings are common with libraries as well
\cite{Kapser_2008_CloneHarmful}. Thus, the uses of similar APIs or
libraries may introduce clones. Coincidentally implementing the same logic by different
developers may also cause cloning. Programmers may unintentionally repeat a common solution for similar kinds of
problems using the common solution pattern of his/her memory to such
similar problems. Therefore, several clones may unknowingly be
created in the software systems.

As can be seen from the above discussion, we need to dig deeper  into each of the root causes so that we can either avoid clones or can keep track of the clones during development making clone management easier in the maintenance as also noted by Zhang et al. \cite{ZhangPXZ12CloningIntent}. We can also think of better programming languages design keeping more abstraction mechanisms for different types of clones at the fingers end of the developers. 

\section{Clone Management Strategies}\label{sec:ManagementStrategy}
For dealing with code clones, Mayrand et al.~\cite{Mayrand_1996_CloneMaintenance} proposed two concrete activities namely ``problem mining" and ``preventive control", which were further supported by a later study of Lague et al.~\cite{Lague_1997_FunctionClone}. Giesecke~\cite{Giesecke_2007_CloneModel} categorized them into compensatory and preventive clone management, respectively.
Giesecke~\cite{Giesecke_2007_CloneModel} suggested that all clone management activities can be associated with one or more of the three categories: corrective, preventive, and compensatory management.

\emph{Corrective clone management} aims for \emph{removal} of existing clones from the system. The objective of \emph{Preventive clone management} is to prevent creation of new clones in the system. \emph{Compensatory clone management} deals with applying techniques (such as annotation, documentation) for  avoiding the negative impacts of clones that are not removed from the system for some valid reasons. In practical settings, avoiding clones may be impossible at times, and the expectation of a clone-free system can be unrealistic. Thus, \emph{preventive} clone management actually refers to \emph{proactive} management~\cite{Hou_2009_CnP,Hou_2009_CloneManage} that aims to deal with the clones during their creation or soon after they are introduced. An opposite strategy, \emph{retroactive clone management}~\cite{Chiu_2007_BeyondDetection} adopts the \emph{post-mortem approach}~\cite{Zibran_2012_CloneSearch}, where clone management activities initiate after the development process is complete up to a milestone. 

Clone management in legacy systems can be the most appropriate for the \emph{post-mortem} strategy. Indeed, prevention is better than cure. Therefore, \emph{proactive} clone management is preferable to \emph{post-mortem} approach. While, ideally, all clones should be managed proactively, in practical settings, proactive treatment for all clones may not be feasible or possible. Therefore, a versatile clone management system should focus on support for proactive management, while at the same time, should also facilitate retroactive clone management~\cite{Chiu_2007_BeyondDetection}. Recently, Zhang et al. \cite{Zhang2013ASE} proposed, \textit{CCEvents} that provides timely notifications about
relevant code cloning events for different stakeholders through continuous monitoring of code repositories. This is one of the first studies on contextual 
and on-demand clone management that clearly shows we need further studies on clone management as well.

\section{Design Space for a Clone Management System}\label{sec:DesignSpace}
Most clone detectors~\cite{ccFinder,NiCad_2008,DECKARD,Higo_2009_Scorpio} are implemented as stand-alone tools separate from IDEs (Integrated Development Environments) and typically search for all clones in a given code base. While clone detection from such tools can help clone management in a  post-mortem approach, researchers and practitioners~\cite{Giesecke_2007_CloneModel,Harder_2010_QuoVadis,Hou_2009_CnP,Hou_2009_CloneManage,Lague_1997_FunctionClone,JSync,Zibran_2011_CloneManage,Zibran_2012_CloneSearch} believe that clone management activities should be integrated with the development process to enable proactive management.

Hou et al.~\cite{Hou_2009_CloneManage}, during the on-going development of their clone management tool \texttt{CnP}~\cite{Hou_2009_CnP}, explored the design space towards tool support for clone management. However, their work was tightly coupled with the clone detection technique based on the programmers' copy-paste operations. Thus, their findings are limited in scope to the management of copy-pasted code, and most of the findings are not applicable to clone management based on similarity based clone detection.

\subsection{Architectural Alternatives of Integration}
\label{sec:arch-altern}

We identify three major alternatives and some sub-alternatives in the design space for a versatile clone management tool. These alternatives are inspired by our experience and the different clone management scenarios reported by Giesecke~\cite{Giesecke_2007_CloneModel}. 

\textbf{Architectural Centrality:}
The need for the integration of clone management activities with the development process suggests that the IDEs should include features to support clone management activities during the actual development phase. While a programmer typically works inside an IDE running on her individual workstation, for fairly large projects, especially in industrial settings, a team of developers collaboratively works on a shared code base kept in a version control system set up on a server. 
Hence, the supports for clone management activities can be implemented as features augmenting the local IDEs, or the functionalities can be implemented at a central repository.

\textbf{Decentralized Architecture:} 
The clone management functionalities, when augmenting the features in local IDEs, can enable the individual programmers to exploit the benefit of clone management. In the decentralized scenario different developers can use different tools, and some programmers can get the flexibility to completely or partially disregard clone management at their respective situations. Apparently, such a decentralized implementation may completely disregard the existence of a central server, and enforces proactive clone management before check-in to the shared repository. However, this necessitates additional requirements for establishing means for communication between distributed developers, as well as combining and synchronizing clone information across all the developers.

\textbf{Centralized Architecture:}
The centralized architecture inherently aims to support clone management in a distributed development process. The functionalities can be implemented as a client-server application on top of central version control systems. Such a centralized clone management system may  require greater effort and offer less flexibility than a decentralized implementation~\cite{Giesecke_2007_CloneModel}. Indeed, a client-server implementation cannot support those individual programmers who work alone on their stand-alone local machines~\cite{Zibran_2012_CloneSearch}. But, the centralized architecture may facilitate the integration of clone detection features with the continuous or periodic (e.g., diurnal) build process \cite{YamanakaCYIS13Notification}.

We currently do not know what strategy works best under which
circumstances. Future research should compare the different ways of
integrating clone management in the development process.

\subsection{Triggering Actors in Clone Management}
A clone management activity may be initiated by the developer or by
the system in response to certain events. 

\textbf{Human Triggered Initiative:} A developer, after writing or modifying a piece of code, may invoke the search for clones in the system, upon finding the clones, she may analyze and decide how to deal with them. In such an ad-hoc triggering scenario, the developer, at times, may forget to perform the necessary clone management. An instance of clone management activity may also be periodically scheduled in advance as part of a larger plan of process activities, and clone management activities can be carried out following the post-mortem approach on the current status of the code base. 

\textbf{System Triggered Initiative:} The development environment can trigger clone management activities in response to certain events, such as saving changes in the code, or check-in of modified code to the central repository having the clone detection capability integrated with the build process. Such events may notify and suggest the developer to perform the required clone management operations. However, care must be taken so that those auto-generated notifications and suggestions do not irritate the developer or hinder her normal flow of work.

\textbf{Scope of Clone Management Activity:}
An instance of clone management activity may be \emph{clone-focused} or \emph{system-focused}. A clone-focused activity deals with a narrow set of clones of a particular code segment of interest. On the contrary, a system-focused clone management activity aims to deal with a broad collection of clones in the entire code base, or particular portions of the system.

We need to further investigate for which kind of events clone
management should be triggered by whom. For changes in clones -- in
particular inconsistent ones -- likely a system should notify a
developer. The challenge for all actions triggered by a system must be
to avoid false alarms. Otherwise developers will soon give up using a
clone management tool. For general quality assurance, a quality
manager might observe trends in clones and take initiatives when she
sees a increase of redundancy. The challenge for human triggered
actions is to provide accurate data on demand and to find significant
indicators of problems.

\begin{figure*}[htp]
\begin{center}
	\includegraphics[scale=0.5]{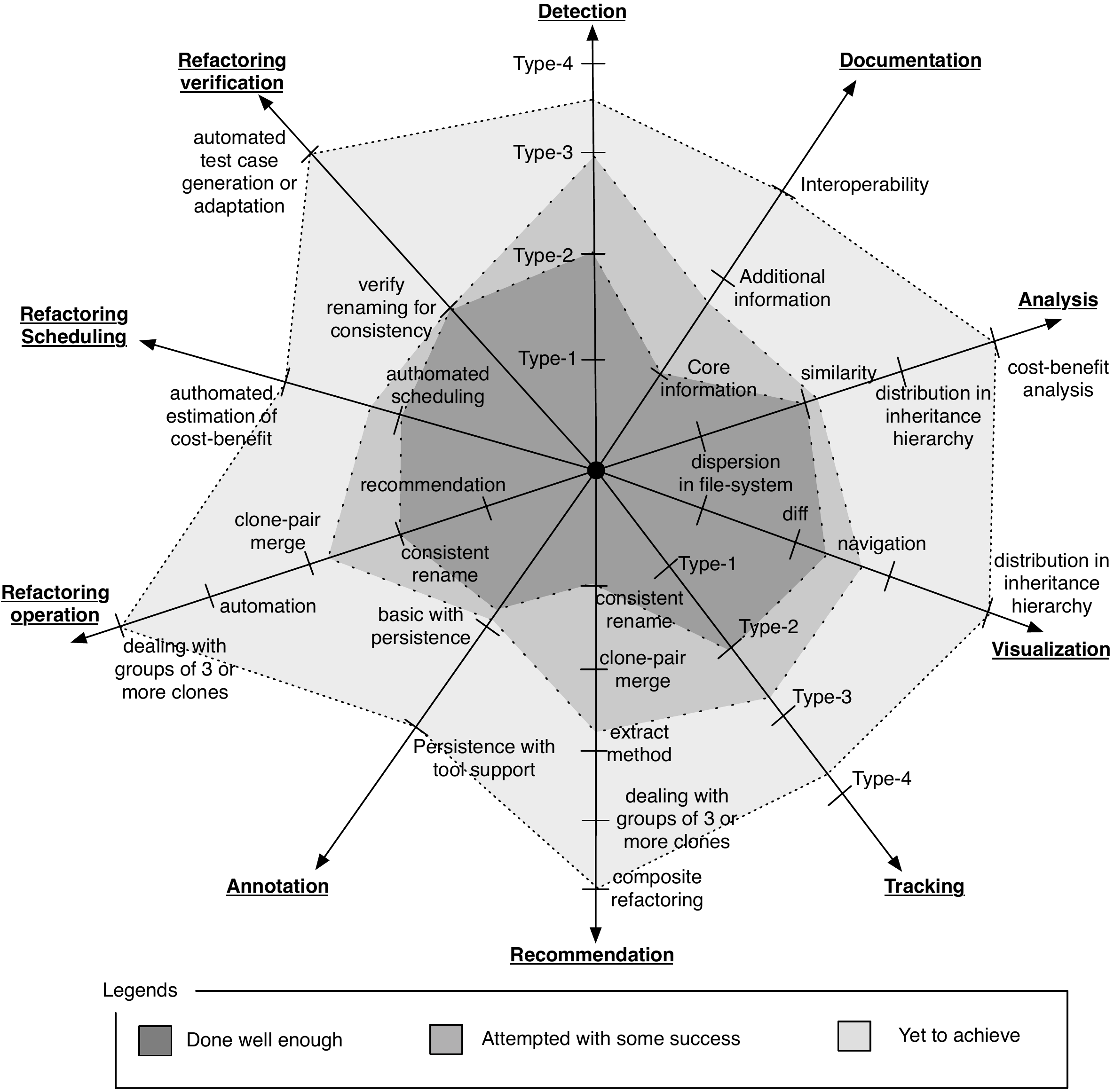}
\caption{Achievements and scopes along different dimensions of clone management activities}
\label{fig:Summary}
\end{center}
\vspace{-0.1cm}
\end{figure*}

\subsection{Scope and Point in Time of Clone Management}

Clones are not restricted to source code. Finding clones in earlier
artifacts may avoid source code clones. 

Clones in requirements documents may lead to duplicate implementation of very similar features if, for example, similar use cases are given to different development teams to implement. In turn, this may result in semantic code clones, or even code segments with very similar structure. Therefore, the detection of clones in requirements specification artifacts could help avoiding clones in code, or to identify semantic clones, which in turn could help to differentiate features from sub-features.

Similarly, clones in design models, especially in model-driven development, may also lead to clones in the source code. Thus, the upfront detection of clones in design models might help to reconsider architectural choices, and result in a leaner, more abstract and essential design resulting in fewer code clones. 

For these reasons, clone management must consider all types of
software artifacts and should be part of early stages in software
development. There are initial studies in detecting clones in other types
of artifacts such as in requirements documents \cite{Juergens_2011_BeyondCode,Juergens_2010_RequirementClone}, in models \cite{Deissenboeck_2008_ModelClone, Storrle_2010_CloneUML}, in sequence diagrams \cite{Liu_2006_SDClone} or in Spreadsheets \cite{HermansSPD13}, which need to explore and deepen further. We envision clone management tools to broaden from
source code to other artifacts and as a consequence a good chance to
avoid source-code clones.

\section{Industrial Adoption of Clone Management}\label{sec:Adoption}
Despite the active research on software clones and their impact on the
development and maintenance of software system, management of code
clones is still far from wide industrial adoption. A reason to this
could be the unavailability of integrated tool support for versatile
clone management. Or maybe industry is just not aware of the
problem. Maybe clones are even not a real problem in the first place
because the advantages outweigh the disadvantages.

What we as researchers need to show first is sufficient empirical
evidence of real problems caused by clones. We have made good progress
in recent years here. Then we need to provide usable working
solutions. We need to demonstrate their benefits in real case
studies. Because benefits are expected to show up only in the long
run, we need long-term studies in realistic industrial settings. Such
long-term industrial studies are difficult to conduct, however.

Despite these difficulties, we see signs that clone management is
gathering momentum in industry. There are several clone detectors
available as Eclipse plug-ins and only recently Microsoft introduced a
clone management feature in Microsoft Visual Studio \cite{DangZGCQX12,
  WangD0ZLM12}. There are several other industrial attempts as well \cite{YamanakaCYIS13Notification, VenkatasubramanyamGS13} 
  including a recent Dagstuhl seminar on the topic \cite{BaxterCCK12Dagstuhl}.

\section{Conclusion}	\label{sec:Conclusion}	

In Figure~\ref{fig:Summary}, we summarize the state of the art along the different dimensions of code clone management and scopes for further improvements. 
Although software clone research matured over the last decade, the majority of the work focused on the detection and analysis of code clones. Compared to those, clone management has earned recent interest due to its practical importance. Notably several surveys~\cite{Koschke_2006_Survey,Pate_2011_EvolutionSurvey,Roy_2007_Survey, Roy_2009_ToolEvaluation, Rattan20131165} appeared in the literature, none of which focused on clone management, and thus a survey on clone management was a timely necessity. This paper presents a comprehensive survey on clone management and pin-points research achievements and scopes for further work towards a versatile clone management system. 

At the fundamental level, the vagueness in the definition of clones at times causes difficulties in formalization, generalization, creation of benchmark data, as well as comparison of techniques and tools. A set of task oriented definitions or taxonomies can address these issues. Most of the integrated tools have limitations in detecting \emph{Type-3} clones, and the detection of \emph{Type-4} clones has still remained an open problem. Moreover, most of the research on software clones so far emphasized clone analysis at different levels of granularity. A variety of techniques for the visualization of clones and the evolution has been proposed. Surprisingly, while clone analysis points to the importance of considering inheritance hierarchy for extracting clone reengineering candidates, there is still not enough visualization support to analyze clones with respect to their existence in the inheritance hierarchy.

Research on clone management beyond detection has mostly been limited
to devising techniques to identify clones. While detection is a
necessity for clone management and many improvements have been
achieved here, filtering and ranking relevant findings is still a
major challenge. It is not yet clear what constitutes a bad clone that
requires treatment. Neither is it sufficiently known what kind of
treatment (refactoring or other types of compensation) works best
under which circumstances. For the bad clones, we need to conduct
root-cause analysis to better understand why they came into existence
and how they could be avoided.

The state of the art demands more research in semi-automated tool support for clone refactoring and cost-benefit analysis of clone removal/refactoring. For integrated clone management, \texttt{JSync}~\cite{JSync} offers a relatively wide set of features compared to others. But, we see that the state of the art is still far from \emph{integrated} tool support, and more is to be done towards a versatile clone management system. Perhaps, due to the unavailability of such tools, there is not much developer-centric ethnographic studies on the patterns of clone management in practice, as well as on the usability and effectiveness of tool support. This survey exposes such potential avenues for further research to create a better impact in the community. 

\textbf{Acknowledgement:} We would like to thank our anonymous reviewers for their useful suggestions and critiques.

\bibliographystyle{plainnat}
\begin{footnotesize}
\bibliography{clones}
\end{footnotesize}
\end{document}